\documentclass[a4paper,10pt]{article}
\usepackage{amssymb}
\usepackage{amsmath}

\def \be {\begin{equation}}
\def \ee {\end{equation}}
\def \ba {\begin{array}{l}}
\def \ea {\end{array}}

\def \bea {\begin{eqnarray}}
\def \eea {\end{eqnarray}}
\def \nn {\nonumber}

\def \a {\alpha}
\def \b {\beta}

\def \d {\delta}

\def \m {\mu}
\def \n {\nu}
\def \k {\kappa}

\def \s {\sigma}
\def \r {\rho}
\def \o {\omega}

\def \th {\theta}
\def \Th {\Theta}

\def \t {\tau}
\def \dag {\dagger}
\def \p {\partial}

\def\bd{\begin{document}}
\def\ed{\end{document}}
\def\nn{\nonumber}
\def\bea{\begin{eqnarray}}
\def\eea{\end{eqnarray}}
\let\bm=\bibitem
\let\la=\label

\def\N{{\cal N}}
\def\sst{\scriptscriptstyle}
\def\thetabar{\bar\theta}
\def\Tr{{\rm Tr}}
\def\one{\mbox{1 \kern-.59em {\rm l}}}

\def\a{\alpha}      \def\da{{\dot\alpha}}
\def\b{\beta}       \def\db{{\dot\beta}}
\def\c{\gamma}  \def\C{\Gamma}  \def\cdt{\dot\gamma}
\def\d{\delta}  \def\D{\Delta}  \def\ddt{\dot\delta}
\def\e{\epsilon}        \def\vare{\varepsilon}
\def\f{\phi}    \def\F{\Phi}    \def\vvf{\f}
\def\h{\eta}
\def\k{\kappa}
\def\l{\lambda} \def\L{\Lambda}
\def\m{\mu} \def\n{\nu}
\def\o{\omega}
\def\P{\Pi}
\def\r{\rho}
\def\s{\sigma}  \def\S{\Sigma}
\def\t{\tau}
\def\th{\theta} \def\Th{\Theta} \def\vth{\vartheta}
\def\X{\Xeta}
\def\z{\zeta}
\def\w{\wedge}
\def\u{\underline}
\def\hs{\hspace}

\def\cA{{\cal A}} \def\cB{{\cal B}} \def\cC{{\cal C}}
\def\cD{{\cal D}} \def\cE{{\cal E}} \def\cF{{\cal F}}
\def\cG{{\cal G}} \def\cH{{\cal H}} \def\cI{{\cal I}}
\def\cJ{{\cal J}} \def\cK{{\cal K}} \def\cL{{\cal L}}
\def\cM{{\cal M}} \def\cN{{\cal N}} \def\cO{{\cal O}}
\def\cP{{\cal P}} \def\cQ{{\cal Q}} \def\cR{{\cal R}}
\def\cS{{\cal S}} \def\cT{{\cal T}} \def\cU{{\cal U}}
\def\cV{{\cal V}} \def\cW{{\cal W}} \def\cX{{\cal X}}
\def\cY{{\cal Y}} \def\cZ{{\cal Z}}
\def\bo {\bar{\o}}

\def\ua{\underline{\alpha}} \def\ubb{\underline{\beta}}
\def\ug{\underline{\gamma}}
\def\ub{\underline{\phantom{\alpha}}\!\!\!\beta}
\def\uc{\underline{\phantom{\alpha}}\!\!\!\gamma}
\def\um{\underline{\mu}} \def\un{\underline{\nu}}
\def\ud{\underline\delta}
\def\ue{\underline\epsilon}
\def\una{\underline a}\def\unA{\underline A}
\def\unb{\underline b}\def\unB{\underline B}
\def\unc{\underline c}\def\unC{\underline C}
\def\und{\underline d}\def\unD{\underline D}
\def\une{\underline e}\def\unE{\underline E}
\def\unf{\underline{\phantom{e}}\!\!\!\! f}\def\unF{\underline F}
\def\unm{\underline m}\def\unM{\underline M}
\def\unn{\underline n}\def\unN{\underline N}
\def\unp{\underline{\phantom{a}}\!\!\! p}\def\unP{\underline P}
\def\unq{\underline{\phantom{a}}\!\!\! q}
\def\unQ{\underline{\phantom{A}}\!\!\!\! Q}
\def\unH{\underline{H}}
\def\ul{\underline}

\def\As {{A \hspace{-6.4pt} \slash}\;}
\def\bs {{b \hspace{-6.4pt} \slash}\;}
\def\Ds {{D \hspace{-6.4pt} \slash}\;}
\def\ds {{\del \hspace{-6.4pt} \slash}\;}
\def\ss {{\s \hspace{-6.4pt} \slash}\;}
\def\ks {{ k \hspace{-6.4pt} \slash}\;}
\def\ps {{p \hspace{-6.4pt} \slash}\;}
\def\pas {{{p_1} \hspace{-6.4pt} \slash}\;}
\def\pbs {{{p_2} \hspace{-6.4pt} \slash}\;}

\def\Fh{\hat{F}}
\def\Vh{\hat{V}}
\def\Xh{\hat{X}}
\def\ah{\hat{a}}
\def\xh{\hat{x}}
\def\yh{\hat{y}}
\def\ph{\hat{p}}
\def\xih{\hat{\xi}}

\def\psit{\tilde{\psi}}
\def\Psit{\tilde{\Psi}}
\def\tht{\tilde{\th}}

\def\At{\tilde{A}}
\def\Qt{\tilde{Q}}
\def\Rt{\tilde{R}}
\def\Nt{\tilde{N}}

\def\at{\tilde{a}}
\def\st{\tilde{s}}
\def\ft{\tilde{f}}
\def\pt{\tilde{p}}
\def\qt{\tilde{q}}
\def\vt{\tilde{v}}
\def\nt{\tilde{n}}

\def\delb{\bar{\partial}}
\def\bz{\bar{z}}
\def\bD{\bar{D}}
\def\bB{\bar{B}}
\def\bo {\bar{\o}}

\def\bk{{\bf k}}
\def\bl{{\bf l}}
\def\bp{{\bf p}}
\def\bq{{\bf q}}
\def\br{{\bf r}}
\def\bx{{\bf x}}
\def\by{{\bf y}}
\def\bR{{\bf R}}
\def\bV{{\bf V}}
\def\bd{\begin{document}}

\def\ed{\end{document}}

\def\d{\delta}\def\D{\Delta}\def\ddt{\dot\delta}

\def\p{\partial} \def\del{\partial}
\def\xx{\times}
\def\uno{\mbox{1 \kern-.59em {\rm l}}}

\def\trp{^{\top}}
\def\inv{^{-1}}
\def\dag{{^{\dagger}}}
\def\pr{\prime}

\def\rar{\rightarrow}
\def\lar{\leftarrow}
\def\lrar{\leftrightarrow}

\def\cw{{\cal W}}
\def\cz{{\cal Z}}
\def\tcm{\tilde{\cal M}}
\def\sgn{{\rm sgn}}
\def\sd {d^{4|4}}
\def\lan{\langle}
\def\ran{\rangle}

\begin{document}

\title{Quasi-normal Modes of Extremal Black Holes from Hidden Conformal Symmetry}
\author{Bin Chen$^{1,2,3}$\footnote{Email:bchen01@pku.edu.cn}\hs{2ex}Jia-ju Zhang$^1$\footnote{Email:jjzhang@pku.edu.cn}\\
\small{$^1$Department of Physics,
and State Key Laboratory of
Nuclear Physics and Technology,}\\
\small{Peking University, Beijing 100871, P.R. China}\\
\small{$^2$Center for High Energy Physics,
Peking University, Beijing 100871, P.R. China}\\
\small{$^3$ Kavli Institute for Theoretical Physics China,
CAS, Beijing 100190, P.R. China}}
\date{\today}
\maketitle

\begin{abstract}
In this paper, we construct the quasi-normal modes of three-dimensional
extremal black holes in an algebraic way. We show that the infinite towers of the
quasi-normal modes of scalar, vector and tensor could be constructed as the descendents of the highest weight modes. Our investigation shows that the hidden conformal symmetry suggested in \cite{Chen:2010fr} is an intrinsic property of the extremal black hole. Moreover, we notice that  we need to fix the freedom in defining the local vector fields and find the right hidden conformal symmetry to obtain the physical quasi-normal modes.
 \end{abstract}

\newpage


\section{Introduction}

In \cite{Castro:2010fd}, it was argued that the hidden conformal symmetry in the
low-frequency scalar scattering off generic non-extremal Kerr black hole was essential
to set up a Kerr/CFT correspondence.
Such hidden conformal symmetry is just a local symmetry rather than a global symmetry, acting on the solution space.  Furthermore in \cite{Castro:2010fd}, a holographic two-dimensional description of generic non-extremal Kerr black hole was proposed. The match of both the entropy counting and the scattering amplitudes provides strong evidence to support this Kerr/CFT conjecture. The similar analysis has been generalized to various cases, see for example \cite{Chen:2010xu,{Chen:2010bh}}.

 Very recently, it was shown in \cite{Chen:2010ik} that the same hidden conformal symmetry may act on the solution space of the vector and tensor as well. The vector fields defining an $SL(2,R)$-algebra act on the tensor via Lie-derivatives. It was shown that for three-dimensional black holes which have two-dimensional conformal field theory (CFT) descriptions the wave equations of the vector and the rank-2 symmetric tensor could be written in terms of  the Lie-induced $SL(2,R)$ quadratic Casimir. This fact shows that the hidden conformal symmetry, if it exists, is an intrinsic property of the black hole, not an artifact of the scalar equation of motion.  Furthermore, the hidden conformal symmetry has been applied to construct the quasi-normal modes (QNM) of the black hole in an algebraic way in \cite{Chen:2010ik}. 
 
 The quasi-normal modes\cite{quasinomral1999} are defined to be the perturbations of the black hole, satisfying the purely ingoing boundary condition at the black hole horizon and appropriate boundary condition at the asymptotical infinity. As a result, the frequencies of the
perturbations are complex, whose imaginary parts determine the decay time of the perturbations.  From AdS/CFT correspondence\cite{{Horowitz:1999jd},Briminghan01}, the quasi-normal modes of the black holes
correspond to the operators perturbing the thermal equilibrium in dual
field theory, and their frequencies are related to the poles  of the retarded correlation function  in the momentum space in the dual conformal field theory. On the other hand, the quasi-normal modes also play an important role in deriving the area law of the black hole entropy in the framework of canonical quantum gravity\cite{Hod:1998vk, Dreyer:2002vy}.  The usual way to determine the quasi-normal spectrum is to solve the equations of motion explicitly under physical boundary condition, whose analytic forms are often out of reach. In \cite{Chen:2010ik}, it was shown that from the hidden conformal symmetry of three-dimensional black holes one could define the highest weight mode, from which the infinite tower of descendent modes could be constructed. This infinite tower of modes turns out to be a tower of quasi-normal modes. More interestingly, the construction works well not only for the scalar, but also  for the vector and the tensor, which requires appropriate composition of the components. The effectiveness of the construction relies on the fact that the hidden conformal symmetry could be defined in the whole spacetime region of three-dimensional black holes rather than just the ``Near" region of the black hole, for various kinds of perturbations via Lie-derivatives.

 In this letter, we generalize the construction proposed in \cite{Chen:2010ik} to the extremal black holes. For the extremal black holes, the discussion of the hidden conformal symmetry is subtle. One needs to introduce a new set of conformal coordinates to define the local $SL(2,R)$ vector fields and the $SL(2,R)$ quadratic Casimir\cite{Chen:2010fr}, which could be used to rewrite the radial equation of the scalar. Even though the conformal coordinate is quite different, one set of hidden conformal symmetry is very similar to the ones in the non-extremal case. This makes the similar construction of the quasi-normal modes feasible. 

In the next section, we give a concise review of the construction in \cite{Chen:2010ik}, paying special attention to several subtle points in the construction. In particular, we find that we need to choose appropriate vector fields to define the physical quasi-normal modes. In Sec. 3, we show how the construction could be generalized to the extremal black holes. In Sec. 4, we take the BTZ black hole and the null warped black hole as explicit examples. For the BTZ black hole, the result we obtain is in perfect match with the one found in the literature. For the null black hole, the construction is slightly different as the equations of motion involve both set of the vector fields. We end with discussions in Sec. 5.

\section{Algebraic construction of QNMs of non-extremal black holes}

In this section we give a brief review of the construction in \cite{Chen:2010ik} and clarify a few subtle points  that were not pointed out explicitly before but are useful in this paper.

\subsection{Hidden conformal symmetry}

The hidden conformal symmetry is not a global symmetry of the black hole. It could only be locally defined and acts just on the solution space. Usually starting from conformal coordinates, one may construct a set of
 vector fields satisfying $SL(2,R)$ algebra
\be \label{ee2} [V_0,V_1]=-V_1,~~[V_0,V_{-1}]=V_{-1},~~[V_1,V_{-1}]=2V_0. \ee
 The general form of the vector fields $V_{0,\pm1}$ for generic non-extremal black holes is of the form
\be \ba \label{z27}
 V_0  = 2A \partial _t  + 2C \partial _\phi  , \\
 V_1  = e^{\mu _1 t + \mu _2 \phi } \left( {\frac{{A\Delta ' + B}}{{\sqrt \Delta  }}\partial _t  + \frac{{C\Delta ' + D}}{{\sqrt \Delta  }}\partial _\phi   + \sqrt \Delta  \partial _r } \right), \\
 V_{ - 1}  = e^{ - \mu _1 t - \mu _2 \phi } \left( {\frac{{A\Delta ' + B}}{{\sqrt \Delta  }}\partial _t  + \frac{{C\Delta ' + D}}{{\sqrt \Delta  }}\partial _\phi   - \sqrt \Delta  \partial _r } \right), \\
\ea \ee
where $\Delta=(r-r_+)(r-r_-)$, $\Delta'=d\Delta/dr$ and the coefficients $\lambda_1,\lambda_2,\mu_1,\mu_2,A,B,C,D$ are constants related to each other \be\label{e1}
  \mu _1  = -\frac{{D}}{{2(AD - BC)}},\hs{3ex}
 \mu _2  = \frac{B}{{2(AD - BC)}}. \ee
It is not hard to see that there are degrees of freedom in the definition of the $SL(2,R)$ vector fields (\ref{z27}). For example, one may redefine
\be \label{z24} \hat V_1=-V_{-1},~~\hat V_0=-V_0,~~\hat V_{-1}=-V_1,  \ee
 such that $\hat V_{0,\pm1}$ still satisfy the $SL(2,R)$ algebra (\ref{ee2}). The redefinition corresponds to change the sign of all the parameters
  $\lambda_1,\lambda_2,\mu_1,\mu_2,A,B,C,D$.   Actually there are more degrees of freedom in defining the $SL(2,R)$ generators. We can define a new set of vector fields as
\be \label{z25} \hat V_1=\alpha V_{1},~~\hat V_0=V_0,~~\hat V_{-1}=\frac{1}{\alpha}V_{-1},\ee
with $\alpha$ as an arbitrary nonzero constant, or
\be \label{z26} \hat V_1=V_{-1},~~\hat V_0=-V_0,~~\hat V_{-1}=V_1. \ee
Both (\ref{z25}) and (\ref{z26}) leave the $SL(2,R)$ algebra (\ref{ee2}) and the Casimir (\ref{e2}) unchanged. In fact the transformation (\ref{z24}) is just using (\ref{z25}) with $\alpha=-1$ and (\ref{z26}) successively. There seems to be no way to fix the above ambiguity in the definition of the vector fields. However from the algebraic construction of the quasi-normal modes, the physical requirement will help us to fix the ambiguity, up to the $\a$ factor which has no physical implication.

For the scalar, the hidden conformal symmetry manifest itself in the fact that the scalar equation of motion could be rewritten in term of the $SL(2,R)$ quadratic Casimir. Moreover, the symmetry could actually be an intrinsic property of the black hole so it acts on the vector and the tensors as well. As the vector fields act on the tensors via Lie-derivative, in order to discuss the hidden conformal symmetry on the vector and the tensor, one may define  the Lie-induced quadratic Casimir  as
\be \label{e2}
\mathcal L^2  =  - \mathcal L_{V_0 } \mathcal L_{V_0 }  + \frac{1}{2}\left( {\mathcal L_{V_1 } \mathcal L_{V_{ - 1} }  + \mathcal L_{V_{ - 1} } L_{V_1 } } \right).
\ee
Acting on a scalar $\Phi$ the Lie-induced Casimir takes the form of
\be \mathcal L^2\Phi=\Pi^{\rho\sigma}\partial_\rho\partial_\sigma\Phi+\Sigma^\rho\partial_\rho\Phi, \ee
where we have defined
\be\label{z10}
\Pi ^{\rho \sigma }  \equiv \frac{1}{2}\left( {V_1^\rho  V_{ - 1}^\sigma   + V_1^\sigma  V_{ - 1}^\rho  } \right) - V_0^\rho   V_0^\sigma,
\ee
\be\label{z11}
\Sigma ^\rho   \equiv \frac{1}{2}\left( {V_1^\sigma  \partial _\sigma  V_{ - 1}^\rho   + V_{ - 1}^\sigma  \partial _\sigma  V_1^\rho  } \right) - V_0^\sigma  \partial _\sigma  V_0^\rho.
\ee
The explicit expressions can be calculated easily, with which we find
\be \label{e5}
 - \mathcal{L}^2\Phi  = \partial _r \Delta \partial _r\Phi  - \frac{\sigma _+ ^2}{{(r - r_ +  )(r_ +   - r_ -  )}}\Phi + \frac{{ \sigma _- ^2 }}{{(r - r_ -  )(r_ +   - r_ -  )}}\Phi.
\ee
with
\be
\sigma_\pm={\left[ {A(r_+-r_-) \pm B} \right]\partial _t  + \left[ {C(r_ +   - r_ -  ) \pm D} \right]\partial _\phi  }.
\ee
As expected, for the scalar, the action of the Lie-induced Casimir is the same as the usual Casimir.

For a vector $A_\mu$, we have
\be
\mathcal L^2 A_\mu   = \Pi ^{\rho \sigma } \partial _\rho  \partial _\sigma  A_\mu   + \Sigma ^\rho  \partial _\rho  A_\mu   + \partial _\mu  \Sigma ^\sigma  A_\sigma   + \Upsilon _\mu ^{\rho \sigma}  \partial _\rho  A_\sigma
\ee
with $\Upsilon _\mu ^{\rho \sigma}$ being defined as
\be
\Upsilon _\mu ^{\rho \sigma }  \equiv V_1^\rho  \partial _\mu  V_{ - 1}^\sigma   + V_{ - 1}^\rho  \partial _\mu  V_1^\sigma   - 2V_0^\rho  \partial _{\mu  } V_0^\sigma.
\ee
Noticing that $\Sigma^\sigma$ is only a function of $r$, if we can find $\kappa_1,\kappa_2$ that satisfy
\be \label{z3}(\kappa_1\partial_t+\kappa_2\partial_\phi)V_{0,\pm1}^\mu=0, \ee
then we will have
\be \mathcal L^2A_+=\Pi^{\rho\sigma}\partial_\rho\partial_\sigma A_++\Sigma^\rho\partial_\rho A_+, \ee
where we have let
\be A_+=\kappa_1 A_t+\kappa_2 A_\phi.\ee
The condition~(\ref{z3}) can be satisfied if
\be \frac{\kappa_1}{\kappa_2}=-\frac{\mu_2}{\mu_1}. \ee
Therefore, under the action of the Lie-induced Casimir  $A_+$ behaves like a scalar. More interestingly, from the equation of motion of a massive vector
\be \epsilon_\mu^{\phantom{\mu}\alpha\beta}\partial_\alpha A_\beta+mA_\mu=0, \ee
we find that for three-dimensional black hole, $A_+$ indeed obeys the relation
\be
(\mathcal{L}^2+m_v^2)A_+=0, \label{lvector2}
\ee
where $m_v$ is a $r$-independent parameter, which may depend on the backgrounds and other quantum number.

For a rank-2 symmetric tensor $T_{\mu\nu}$, the action of the Lie-induced Casimir takes
a more involved way. Nevertheless, the following combination of the components
\be T_ +   = \kappa _1 T_{tt}  + \kappa _2 T_{t\phi }  + \kappa _3 T_{\phi t}  + \kappa _4 T_{\phi \phi } \ee
 behaves like a scalar
\be \mathcal L^2T_+=\Pi^{\rho\sigma}\partial_\rho\partial_\sigma T_++\Sigma^\rho\partial_\rho T_+, \ee
if $\kappa_1,\kappa_2,\kappa_3,\kappa_4$ are chosen to satisfy
\be \frac{\kappa_1}{\kappa_2}=\frac{\kappa_3}{\kappa_4}=-\frac{\mu_2}{\mu_1}, ~~\kappa_2=\kappa_3.  \ee
Similarly, from the equation of motion of $T_{\mu\nu}$ in three-dimensional spacetime
\be \epsilon_\mu^{\phantom{\mu}\alpha\beta}\nabla_\alpha T_{\beta\nu} +mT_{\mu\nu}=0, \ee
we find that
\be
(\mathcal{L}^2+m_t^2)T_{+}=0, \label{ltensor}
\ee
where $m_t$ is a $r$-independent parameter, which could depend on the backgrounds and other quantum number.

In short, with the Lie-induced quadratic Casimir (\ref{e2}), we can discuss all kinds of perturbations in an uniform way. With appropriate combination, their equations of motion
could all be put into the form
\be\label{uniform}
(\mathcal{L}^2+m_{\hat t}^2)\hat T_+=0, \ee
where $\hat T_+$ stands for a general perturbation and $m^2_{\hat t}$ is its corresponding $r$-independent parameter.
It is remarkable that $m^2_{\hat t}$ may not just the mass parameter, it could depend on other quantum numbers. In particular, for the three-dimensional warped black holes, $m^2_{\hat t}$ can include the generators of the other set of $SL(2,R)$ vector fields.

For non-extremal black holes, there exist another set of vector fields forming hidden conformal symmetry. The vector fields take a quite similar form as (\ref{z27}) with a different set of parameters  $\bar\lambda_1,\bar\lambda_2,\bar\mu_1,\bar\mu_2,\bar A,\bar B,\bar C,\bar D$ satisfying a similar relationship as (\ref{e1}). And we  have also similar Lie-induced Casimir $\mathcal{\bar L}^2$  (\ref{e2}) and (\ref{e5}).


\subsection{Constructions of quasi-normal modes}

The construction of various kinds of quasi-normal modes is similar.
Let us take the scalar as the example to illustrate the construction. For the scalar we  have
\be \left(\mathcal{L}^2+m_s^2\right)\Phi=0. \ee
Firstly, we define the highest weight state:
\be \label{hweight}
\mathcal{L}_{V_0}\Phi^{(0)}=h_R\Phi^{(0)}, ~~\mathcal{L}_{V_1}\Phi^{(0)}=0,
\ee
from which the infinite tower of quasi-normal modes can be constructed as its descendent
\be \label{e11}
\Phi^{(n)}=\left(\mathcal{L}_{V_{-1}}\right)^n\Phi^{(0)}, \hs{3ex} n\in N.
\ee
The scalar field $\Phi$ could be expanded as
\be \Phi=e^{-i\omega t+ik\phi}R(r), \ee
where $\o, k$ are quantum numbers corresponding to the Killing vector $\p_t, \p_\phi$ of the black hole background. As
\be
[\mathcal{L}_X,\mathcal{L}_Y]=\mathcal{L}_{[X,Y]},\hs{3ex}\mathcal{L}_{aX}=a\mathcal{L}_X
\label{lie} \ee where $X,Y$ are arbitrary vectors and $a$ is an
arbitrary constant,  the Lie-induced Casimir always commute with the Lie-derivatives $\mathcal{L}_{V_i}$.  The highest weight condition (\ref{hweight})
 gives the conformal weight
\be \label{z1} h_R=\frac{1}{2}\left(1+\sqrt{1+4m_s^2}\right). \ee
And the frequencies of the quasi-normal modes could be read from (\ref{e11})
\be \label{z2} 2A\omega_R^{(n)}=2Ck+i(h_R+n), \ee
with $n$ being a non-negative integer. For the
highest-weight mode $\Phi^{(0)}$, we have
\begin{equation}
2A\omega_0-2C k_0=ih_R,
\end{equation}
where $\omega_0$ and $k_0$ are its frequency and  angular momentum.
In principle,  $k_0$ could be complex in the solution. Taking the
highest mode as quasinormal modes require $k_0$ be real. For the
descendent mode $\Phi^{(n)}$, we have \be
\mathcal{L}_{V_0}\Phi^{(n)}=(-i2A\omega_R^{(n)}+i2C
k^{(n)})\Phi^{(n)}, \ee where its frequency $\omega_R^{(n)}$ and angular
momentum $k_R^{(n)}$ are related to $\o_0$ and $k_0$ via the
relation
\begin{equation}
\omega_R^{(n)}=\omega_0-in\mu_1,\hs{5ex}
k_R^{(n)}=k_0+in\mu_2.\label{relationRk}
\end{equation}
To be a well-defined quasinormal mode, the angular momentum
$k_R^{(n)}$ should be real, which requires a choice of complex
$k_0$. Note that the real part of the $k_0$ and $k_R^{(n)}$ are
always the same, taken as $k$. From the relation (\ref{relationRk})
  we obtain  (\ref{z2}) as well. For the vector and the tensor, the frequencies of the quasi-normal modes take the same form as (\ref{z2}) with different conformal weights.

As the quasi-normal modes are defined as the modes decaying with the time, the imaginary part of their frequencies should be negative. This indicates that in (\ref{z2}) $A$ should be negative.
In the non-extremal cases we can always use the degree of freedom in defining the $SL(2,R)$ vector fields and choose the sign of $A$ to satisfy the $A<0$ condition. However in the next section we will see that in the extremal cases the freedom in choosing the sign does not exist, then some tricks have to be used to construct the quasi-normal modes. If $A>0$, from (\ref{z26})  we have
\be \hat V_0=2\hat A \partial_t+ 2\hat C \partial_\phi. \ee
with the desiring condition $\hat A=-A<0$ and $\hat C=-C$. Using $\hat V_{0,\pm1}$ we can construct a tower of quasi-normal modes,
\be \label{ee42}
\mathcal{L}_{\hat V_0}\Phi^{(0)}=h_R\Phi^{(0)}, ~~\mathcal{L}_{\hat V_1}\Phi^{(0)}=0,~~
\Phi^{(n)}=\left(\mathcal{L}_{\hat V_{-1}}\right)^n\Phi^{(0)}.
\ee
 From this construction,  the conformal weight (\ref{z1}) is unchanged, while the frequencies of the quasi-normal modes are changed to
\be 2A\omega_R^{(n)}=2Ck-i(h_R+n), \ee
which have the desired property of  quasi-normal modes.

With the quasi-normal modes defined as~(\ref{e11}), we can solve the equation of the highest weight mode $\mathcal{L}_{V_1}\Phi^{(0)}=0$ explicitly, and find the solution to be
\be
\Phi ^{(0)}  = C_0 (r - r_ +  )^{iA\omega^{(0)}_R  - iCk + \frac{{iB\omega^{(0)}_R  - iDk}}{{r_ +   - r_ -  }}} (r - r_ -  )^{iA\omega^{(0)}_R  - iCk - \frac{{iB\omega^{(0)}_R  - iDk}}{{r_ +   - r_ -  }}} e^{ - i\omega^{(0)}_R t + ik\phi },
\ee
where $C_0$ is an integration constant. In the limit $r \to r_+$,
\be
\Phi ^{(0)}  \propto {\mathop{\rm e}\nolimits} ^{i\omega^{(0)}_R \left[ {\left( {A + \frac{B}{{r_ +   - r_ -  }}} \right)\ln (r - r_ +  ) - t} \right]}.
\ee
To satisfy the ingoing boundary condition at the horizon $r=r_+$, we need
\be \label{z5} A+\frac{B}{r_+-r_-}<0. \ee
In all the cases we studied, we always have $AB\geq 0$. Therefore, the requirement that the quasi-normal modes are decaying is consistent with the ingoing boundary condition.

 Noting that
\be V_{-1}\propto(r-r_+)^{-\frac{1}{2}}e^{-\mu_1 t},~~\mbox{as} ~~r\to r_+, \ee
we have
\be
\Phi ^{(n)}  \propto {\mathop{\rm e}\nolimits} ^{i\omega _R^{(n)} \left[ {\left( {A + \frac{B}{{r_ +   - r_ -  }}} \right)\ln (r - r_ +  ) - t} \right]} (r - r_ +  )^{ - n\beta } ,~~n \in N,
\ee
where
\be
 \omega _R^{(n)}  = \omega _R^{(0)}  - in\mu _1 , \hs{3ex}
 \beta  = \mu _1 \left( {A + \frac{B}{{r_ +   - r_ -  }}} \right) + \frac{1}{2}.
\ee
 This shows that $\Phi^{(n)}$ satisfy the ingoing boundary condition as well. Also in concrete examples we always have $\mu_1\geq0$, so if we have ${\rm Im} ~\omega_R^{(0)} \le 0$, then ${\rm Im} ~\omega_R^{(n)} \le 0$ for every $n\in N$.

In the $r\to\infty$ limit, we have asymptotically
\be
\Phi^{(0)}\propto r^{i(2A\omega_R^{(0)}-2Ck)}=r^{-h_R}.
\ee
As asymptotically
\be
 V_{-1} \sim e^{-\mu_1t-\mu_2\phi}(2A\partial_t+2C\partial_\phi-r\partial_r),
\ee
 for arbitrary $n\in N$ we always have
\be
\Phi^{(n)}\propto r^{-h_R}.
\ee
So we can see that the solution has the right behavior as the quasi-normal modes.

If we get the vector fields with $A>0$, we can either add an overall minus sign for all the coefficient, which equals the transformation (\ref{z24}), or use the transformation (\ref{z26}), and then construct well-defined quasi-normal modes as the $A<0$ case.

\section{Quasi-normal modes of extremal black holes}



   The hidden conformal symmetry of extremal black holes can not be simply obtained by taking the limit $r_-\to r_+$ from that of the non-extremal ones. With the conformal coordinates  proposed in \cite{Chen:2010fr}, we can define  two sets of vector fields both satisfying $SL(2,R)$-algebra
\be\ba
 V_1  = \frac{2}{A_1}(2\pi T_L \partial _t  - 2n_L \partial _\phi  ), \\
 V_0  =  - (r - r_ +  )\partial _r  + \frac{1}{A_1}(\alpha _1 t + \beta _1 \phi )(2\pi T_L \partial _t  - 2n_L \partial _\phi  ), \\
 V_{ - 1}  =  - (\alpha _1 t + \beta _1 \phi )(r - r_ +  )\partial _r  + \frac{{\gamma _1 }}{{A_1(r - r_ +  )}}(\beta _1 \partial _t  - \alpha _1 \partial _\phi  ) \\
 \begin{array}{*{20}c}
   {} & {}  \\
\end{array} + \left[ {(\alpha _1 t + \beta _1 \phi )^2  + \left( {\frac{{\gamma _1 }}{{r - r_ +  }}} \right)^2 } \right]\frac{1}{{2A_1}}(2\pi T_L \partial _t  - 2n_L \partial _\phi  ) ,\\
\ea \ee
\be\label{z8}
\begin{array}{l}
 \tilde V_1  = 2e^{ - 2\pi T_L \phi  - 2n_L t} \left[ {(r - r_ +  )\partial _r  - \frac{1}{{A_1}}(\beta _1 \partial _t  - \alpha _1 \partial _\phi  ) - \frac{{\gamma _1 }}{{A_1(r - r_ +  )}}(2\pi T_L \partial _t  - 2n_L \partial _\phi  )} \right], \\
 \tilde V_0  =  - \frac{1}{{A_1}}(\beta _1 \partial _t  - \alpha _1 \partial _\phi  ) ,\\
 \tilde V_{ - 1}  =  - \frac{1}{2}e^{2\pi T_L \phi  + 2n_L t} \left[ {(r - r_ +  )\partial _r  + \frac{1}{{A_1}}(\beta _1 \partial _t  - \alpha _1 \partial _\phi  ) + \frac{{\gamma _1 }}{{A_1(r - r_ +  )}}(2\pi T_L \partial _t  - 2n_L \partial _\phi  )} \right].
 \end{array}
\ee
 Since the Hawking temperature of an extremal black hole is vanishing,  the temperature in one sector of the dual  conformal field theory must be vanishing, while the one in the other sector could be non-zero, which is labeled as $T_L$ in the above relations.

 Obviously $\{V_{0,\pm1}\}$ looks quite different from the one (\ref{z27}) we have been discussing,
 but $\{\tilde V_{0,\pm1}\}$ is quite similar to (\ref{z27}). In fact, using the transformation (\ref{z25}) with $\alpha=1/2$ we have
 exactly the extremal form of (\ref{z27}). This fact allows us to construct the quasi-normal modes of extremal black holes in the similar way as shown in last section.

To get the action of the Lie-induced Casimir from $\tilde V_{0,\pm1}$ on the scalar  we have to calculate $\Pi ^{\rho \sigma }$, $\Sigma ^\rho$ in (\ref{z10}), (\ref{z11}) explicitly using (\ref{z8}).
Then we have
\be \nn
-\mathcal{L}^2\Phi  = \partial_r\Delta \partial_r\Phi - \frac{{\gamma _1 ^2 (2\pi T_L \partial _t  - 2n_L \partial _\phi  )^2 }}{{A_1 ^2 (r - r_ +  )^2 }}\Phi - \frac{{\gamma _1 (2\pi T_L \partial _t  - 2n_L \partial _\phi  )(2\beta _1 \partial _t  - 2\alpha _1 \partial _\phi  )}}{{A_1 ^2 \left( {r - r_ +  } \right)}}\Phi,
\ee
with $A_1=2\pi T_L\alpha_1 - 2n_L \beta_1$. 

Similarly we can discuss the action of Lie-induced $SL(2,R)$ quadratic Casimir on the vector and the symmetric rank-2 tensor. We see in (\ref{z13}) that $\Sigma^\sigma$ has the same form as the non-extremal cases, and only depends on $r$. There are $\tilde\kappa_1,\tilde\kappa_2$ with
\be \label{eq12}
 \frac{{\tilde \kappa _1 }}{{\tilde \kappa _2 }} =  - \frac{{\pi T_L }}{{n_L }}
\ee
satisfying
\be (\tilde \kappa _1 \partial _t  + \tilde \kappa _2 \partial _\phi  )\tilde V_{0, \pm 1}^\mu   = 0,\ee
then we can make appropriate combination of the components of the vector or the tensor such that after appropriate combination the vector and the tensor both behave as a scalar under the action. Within our expectation, the equations of motion of the perturbations could be cast into an uniform form as (\ref{uniform}).


The construction of the quasi-normal modes follows straightforwardly. Let us just take the scalar as an example. For the scalar we have
\be \label{z22}
\left( \mathcal{L}^2  + m_s^2  \right)\Phi  = 0,
\ee
then we can define the quasi-normal modes as
\be \label{z15}
\mathcal{L}_{\tilde V_0 }  \Phi ^{(0)}  = h_L \Phi ^{(0)} ,~~\mathcal{L}_{\tilde V_1 } \Phi ^{(0)}  = 0,~~\Phi ^{(n)} = \left( \mathcal{L}_{\tilde V_{-1}} \right)^n \Phi ^{(0)} ,~~n \in N.
\ee
This determines the conformal weight and the frequencies of the quasi-normal modes
\be \label{z12}h_L  = \frac{1}{2}\left( {1 + \sqrt {1 + 4 m_s^2} } \right), \ee
\be \label{z20}\omega _L^{(n)}  =  - \frac{{\alpha _1 }}{{\beta _1 }}k - i\frac{{A_1 }}{{\beta _1 }}(h_L  + n), \hs{3ex} n\in N. \ee

From the previous discussion we can see the method only applies to the case
$\frac{\beta_1}{A_1}>0$.
For the case $
\frac{\beta_1}{A_1}<0$,
we have to redefine the vector fields as  (\ref{z26}). Then we  get the same conformal weight  as (\ref{z12}) and the desired frequencies of the quasi-normal modes
\be \label{z21}\omega _L^{(n)}  =  - \frac{{\alpha _1 }}{{\beta _1 }}k +i\frac{{A_1 }}{{\beta _1 }}(h_L  + n). \ee


With the quasi-normal modes constructed as (\ref{z15}), we get the solution to $\mathcal{L}_{\tilde V_1 } \Phi ^{(0)}  = 0$ as
\be
\Phi ^{(0)}  = C_0 \left( {r - r_ +  } \right)^{ - \frac{{i(\beta _1 \omega_L^{(0)}  + \alpha _1 k)}}{{A_1 }}} e^{\frac{{i\gamma _1 (2\pi T_L \omega_L^{(0)}  + 2n_L k)}}{{A_1 (r - r_ +  )}} - i\omega_L^{(0)} t + ik\phi }.
\ee
In the limit $r \to r_+$,
\be
\Phi ^{(0)}  \propto {\rm{e}}^{i\omega_L^{(0)} \left[ {\frac{{\gamma _1 2\pi T_L }}{{A_1 (r - r_ +  )}} - t} \right]}.
\ee
To satisfy the ingoing boundary condition at the horizon $r=r_+$, we need
\be \label{z17}\frac{{\gamma _1 }}{{A_1 }} > 0. \ee
Similar to the non-extremal cases, we have no general proof that (\ref{z16}) and (\ref{z17}) can be satisfied simultaneously, which however turns out to be true for the cases we study in this paper. Similarly,
 $\Phi^{(n)}$ also satisfy the ingoing boundary condition.

In the $r\to\infty$ limit, we have asymptotically
\be
\Phi^{(0)}\propto r^{ - \frac{{i(\beta _1 \omega_L^{(0)}  + \alpha _1 k)}}{{A_1 }}}=r^{-h_L},
\ee
\be
 \tilde V_{-1}  \sim -\frac{1}{2}e^{2\pi T_L \phi  + 2n_L t} \left[ r\partial_r+\frac{1}{A_1}(\beta _1 \partial _t  - \alpha _1 \partial _\phi ) \right],
\ee
and so for an arbitrary $n\in N$ we always have
\be
\Phi^{(n)}\propto r^{-h_L}.
\ee
Therefore we see that the tower of descendents we constructed have the right behavior as the quasi-normal modes.

\section{Examples}

In this section, we illustrate our construction with two concrete examples. One example is extremal BTZ black hole, whose quasi-normal modes have been  studied in \cite{Crisostomo:2004hj}. The other one is null warped black hole, whose quasi-normal modes have been discussed in \cite{ChenXu2}.

\subsection{BTZ black hole}

The BTZ black hole is a vacuum solution for (2+1)-dimensional Einstein gravity with a negative cosmological constant. The metric of the BTZ black hole is\cite{BTZ}
\be
ds^2  =  - \frac{{(r^2  - r_ + ^2 )(r^2  - r_ - ^2 )}}{{r^2 }}dt^2  + \frac{{r^2 }}{{(r^2  - r_ + ^2 )(r^2  - r_ - ^2 )}}dr^2  + r^2 \left( {d\phi  - \frac{{r_ +  r_ -  }}{{r^2 }}dt} \right)^2.
\ee
This black hole could be  described holographically in terms of 2D CFT with the left and right moving temperature
\be T_L=\frac{r_++r_-}{2\pi},~~T_R=\frac{r_+-r_-}{2\pi}. \ee
The quasi-normal modes of BTZ black hole have been widely studied in the literature, for an incomplete list 
see \cite{BTZ}. 

The hidden conformal symmetry of extremal BTZ black hole has not been discussed in \cite{Chen:2010fr}, but the discussion follows straightforwardly.   In this case, we should just replace $r$ with $r^2$ and $\partial_r$ with $\partial_{r^2}$ in the conformal coordinates and the vector fields. The equation of motion of a scalar is
\be \label{z23}
\partial _{r^2 } (r^2  - r_ + ^2 )^2 \partial _{r^2 } \Phi - \frac{{r_{^ +  }^2 (\partial _t  + \partial _\phi  )^2 }}{{4(r^2  - r_ + ^2 )^2 }}\Phi- \frac{{(\partial _t  + \partial _\phi  )(\partial _t  - \partial _\phi  )}}{{4(r^2  - r_ + ^2 )}}\Phi  = \frac{{m^2 }}{4}\Phi,
\ee
and the coefficients in (\ref{ccord}) turn out to be
\be
 \alpha _1  = \beta _1  =  \frac{{\gamma _1 }}{{r_ +  }}, \hs{3ex}
 T_L  = \frac{{r_ +  }}{\pi }, \hs{3ex}
 n_L  =  - r_ + .
 \ee
 From (\ref{eq12})  we have $\tilde\kappa_1/\tilde\kappa_2=1$ , so we have the combination \be
 A_+=A_t+A_\phi, \ee and
 \be
 h_+=h_{tt}+h_{t\phi}+h_{\phi t}+h_{\phi\phi},
 \ee
both of which behave like a scalar under the Lie-induced Casimir. Actually $A_+$ and $T_+$ are exactly the same ones found in the non-extremal case\cite{Chen:2010ik}.

The equations of motion of the scalar, the vector and the tensor could all be put into the form
\be\label{uniform}
(\mathcal{L}^2+m_{\hat t}^2)\hat T_+=0, \ee
where
\be
\mathcal{L}^2  = 4\partial _{r^2 } (r^2  - r_ + ^2 )^2 \partial _{r^2 }  - \frac{{r_{^ +  }^2 (\partial _t  + \partial _\phi  )^2 }}{{(r^2  - r_ + ^2 )^2 }} - \frac{{(\partial _t  + \partial _\phi  )(\partial _t  - \partial _\phi  )}}{{r^2  - r_ + ^2 }},
\ee
and $m_{\hat t}^2$ take different values for different perturbations
\be
m_s^2=\frac{1}{4}m^2,\hs{3ex}
m_v^2=\frac{1}{4}(m^2+2m),
\hs{3ex}m_t^2=\frac{1}{4}(m^2+4m+3).
 \ee
As
\be \frac{\beta_1}{A_1}=\frac{1}{4r_+}>0, \ee
we may define the quasi-normal modes as (\ref{z15}). Using (\ref{z12}), (\ref{z20}) we finally obtain
\bea
h_L^s=\frac{1}{2}\left(1+\sqrt{1+m^2}\right),\hs{3ex}
h_L^v=\frac{m}{2}+1,\hs{3ex}
h_L^t=\frac{m+3}{2},
\eea
and the frequencies of the quasi-normal modes which are of the same form for all perturbations
\be \omega_L^{(n)}=-k-i4\pi T_L(h_L+n),  \hs{3ex}n\in N. \ee
 This is in perfect match with the one found  in \cite{Chen:2010ik}\footnote{A similar construction on tensor quasi-normal modes of extremal BTZ black hole in 3D TMG has been presented in \cite{Afshar:2010ii}. Our result is in consistent with theirs.}.
 
In \cite{Crisostomo:2004hj}, it has been shown that the quasi-normal modes of extreme BTZ black hole are absent, due to the vanishing Hawking temperature in this case. Our study gives different results. The BTZ black hole could be holographically described by a two-dimensional CFT with both left and right-moving temperatures\cite{Briminghan01}. For extreme BTZ, the fact that its Hawking temperature is vanishing suggests that one sector of the dual CFT is frozen to zero temperature, but the other sector is still active with a nonvanishing temperature. In this case, the quasi-normal modes still correspond to the the operators perturbing the thermal equilibrium in dual chiral CFT. Our results fit nicely with this picture. 

\subsection{Null warped $AdS_3$ black hole}

The null warped $AdS_3$ spacetime is one of the vacuum solution of (2+1)-dimensional topological massive gravity. The null warped black hole could be taken as the quotient of the null warped $AdS_3$. The metric of the null warped $AdS_3$ black hole is of the form\cite{Andy08}
\be ds^2=-2rdtd\phi+\frac{dr^2}{4r^2}+(r^2+r+\alpha^2)d\phi^2, \ee
where $1/2>\alpha>0$ in order to avoid the naked causal singularity. The horizon is located at $r=0$. The null warped black hole is an extremal black hole. In \cite{ChenXu2,{Anninos:2010pm}}, it was conjectured that the null warped black hole could have a holographic description in terms of a two-dimensional chiral conformal field theory with central charge and the temperature\footnote{The chiral CFT has been suggested to have central charge and temperature in the right sector. Here to be in match with our notation, we re-label them as in left sector}
\be C_L=2l/G, \hs{3ex}
 T_L=\frac{\a}{\pi l}.
 \ee

The hidden conformal symmetry of this black hole has been addressed in \cite{Chen:2010fr}. The equation of motion for the scalar field with mass $m$ takes the form
\be \label{eq15}
\partial_rr^2\partial_r\Phi-\frac{\alpha^2\partial_t^2}{4r^2}\Phi-\frac{\partial_t(\partial_t+2 \partial_\phi)}{4r}\Phi-\frac{\partial_t^2}{4}\Phi=\frac{m^2}{4}\Phi,
\ee
which allows us to read out the parameters in the conformal coordinate
\be \alpha_1=-2\beta_1,~~\gamma_1=\alpha \beta_1,~~T_L=\frac{\alpha}{\pi},~~n_L=0.\ee
As $V_1$, $\tilde V _0$ could be given explicitly as
\be  \label{eq14}
V_1=-\frac{1}{\beta_1}\partial_t,\hs{3ex}
\tilde V_0=\frac{1}{4\alpha}(\partial_t+2\partial_\phi),
 \ee
the scalar equation could be rewritten as
\be (\mathcal{\tilde L}^2+b\mathcal{L}_{V_1}^2+m_s^2)\Phi=0, \label{nullscalar}\ee
where $b=\b_1^2/4, m_s^2=m^2/4$ and $\mathcal{L}_{V_1}\Phi^{(0)}=q\Phi^{(0)}$ with $q=i\frac{\o}{\b_1}$. Since
\be
\frac{\beta_1}{A_1}=-\frac{1}{4\alpha}<0,
\ee
we have to redefine the vector fields as (\ref{z26}) and then construct the quasi-normal modes. From (\ref{z12}) we find the conformal weight
\be h_L=\frac{1}{2}\left( {1 + \sqrt {1 + 4(m_s^2+bq^2)} } \right), \ee
and the frequencies of the quasi-normal modes
\be \omega _L^{(n)}  = 2k - i4 \pi T_L(n+h_L), \hs{3ex}n\in N. \ee
In the scalar case we can easily identify $(m_s^2+bq^2)$ with $(m^2-\omega^2)/4$. Taking into account of the following identification between quantum numbers\cite{ChenXu2,Chen:2009cg},
\be k =  - \tilde \omega, ~~w = 2\tilde k, \label{quan}\ee
we finally obtain
\be \ba \label{eq17}
 h_L  = \frac{1}{2}\left( {1 + \sqrt {1 + m^2  - 4\tilde k^2 } } \right), \\
 \tilde \omega _L^{(n)}  =  - \tilde k - i2\pi T_L (n+h_L), \hs{3ex}n\in N. \\
\ea \ee
The result is in perfect match with~\cite{ChenXu2}.

Using (\ref{eq12}) and $n_L=0$ we can choose $\tilde\kappa_1=1,\tilde\kappa_2=0$. As a result, in the linear combination of the vector and the tensor, only $A_t$ and $h_{tt}$ appear. In other words, the vector component $A_t$ and tensor component $h_{tt}$ can be treated  as the scalar under the action of the Lie-induced Casimir. Their equations of motion are the same as (\ref{nullscalar}) but with different mass paramaters
\bea
m^2_v = \frac{1}{4}(m^2-2m) \hs{3ex}
m^2_t = \frac{1}{4}(m^2-4m+3).
\eea

The quasi-normal modes of the vector and the tensor are of the same form as (\ref{eq17}), but the conformal weights are a little different
\begin{equation}
h_L^v  = \frac{1}{2}\left( {1 + \sqrt {1 + m^2-2m-4\tilde k^2 } } \right), \hs{3ex}
h_L^t  = \frac{1}{2}\left( {1 + \sqrt {1 + m^2-4m+3- 4\tilde k^2 } } \right).
\end{equation}
Here we have taken into account of the identification (\ref{quan}).
$h_L^v$ is the same as the one found in~\cite{ChenXu2}. The conformal weight of the massive symmetric tensor mode has not been addressed before. In the paper~\cite{Anninos:2010pm}, the gravitational perturbation has been investigated. It was found to  have the conformal weight
\be
h^t=\frac{1}{2}\left( {1 + \sqrt {1 - 4\tilde k^2 } } \right), \label{nullweight}
\ee
which corresponds to  $m=3$ or $m=1$ in our case. Recall that in three-dimensional topological massive gravity, the gravity perturbation could be massive. For null warped spacetime, it is only well-defined at $\nu=1$  which may give a massive graviton with mass $m=3$. In this sense, our result is compatible with (\ref{nullweight}).

\section{Discussions}

In this paper, we studied the quasi-normal modes of three-dimensional extremal black holes, which have dual two-dimensional holographic CFT descriptions and therefore hidden conformal symmetries. As the non-extremal case, the quasi-normal modes could be constructed in an algebraic way. The towers of quasi-normal modes were constructed as the descendent modes of the highest weight mode. Different from the non-extremal case, there is only non-vanishing temperature in one sector of the dual 2D CFT of extremal black hole such that there is at most one set of quasi-normal modes. This is in accordance with the fact that only one set of $SL(2,R)$ vector fields of the extremal black hole could be applied to construct the quasi-normal modes.

For the warped black holes in 3D TMG, things get more interesting. In fact, even for the non-extremal black hole, the scalar Laplacian generically could not be written purely in terms of a simple $SL(2,R)$ quadratic Casimir. The Lie-derivative of a vector field must be included in the parameter $m^2_{\hat t}$, even though such term does not induce $r$-dependence. As a result, only one set of quasi-normal modes could be constructed\cite{Chen:2010ik}. In the extremal limit,  this set of quasi-normal modes may fail to survive. This happens for the spacelike stretched warped AdS$_3$ black hole and self-dual warped black hole\cite{Bin10selfdual}.  On the other hand, in these two cases, there is really no quasi-normal mode from the point of view of  retarded Green's functions, which have the poles in the quantum number $k$, but no pole in the frequency.  We do not know if this is just a coincidence or there is a deep reason behind it.

Another interesting point is that the construction of the quasi-normal modes may help us to fix the ambiguity in defining the hidden conformal symmetry. As we showed in Section 2,  there exist degrees of freedom in defining the local vector fields, which may satisfy the $SL(2,R)$ algebra and lead to the same $SL(2,R)$ quadratic Casimir. Naively one may construct the quasi-normal modes with any set of these vector fields. However, not of all of them lead to physically acceptable quasi-normal modes. The decaying nature of the quasi-normal modes picks out one set of the vector fields, which we believe to be the right hidden conformal symmetry of the black hole. The other choice of vector fields may lead to a set of modes of enhancing nature, which is physically unacceptable. As the decaying nature of the quasi-normal modes is equivalent to the ingoing boundary condition at the horizon, which is required to compute the retarded Green's function from AdS/CFT prescription\cite{Son:2002sd}, the time-growing modes correspond to the poles of advanced Green's function, or equivalently imposing the outgoing boundary condition at the horizon. Therefore,  the choice on the vector fields is closely related to the choice of the boundary conditions at the horizon.

\section*{Acknowledgments}

The work was in part supported by NSFC Grant No. 10775002, 10975005.
We would like to thank Jiang Long for valuable discussions.


\begin{thebibliography}{99}
\bibitem{Castro:2010fd}
 A.~Castro, A.~Maloney and A.~Strominger,
  Phys. Rev. D{\bf 82}:024008,2010 [arXiv:1004.0996[hep-th]].

 \bibitem{Chen:2010xu}
  B.~Chen and J.~Long,
  JHEP {\bf 1006}, 018 (2010)
  [arXiv:1004.5039 [hep-th]].

\bibitem{Chen:2010bh}
  B.~Chen and J.~Long,
  JHEP {\bf 1008}, 065 (2010)   [arXiv:1006.0157 [hep-th]].

\bibitem{Chen:2010ik}B.~Chen and J.~Long, Phys. Rev. D{\bf 82}:126013,2010 [arXiv:1009.1010[hep-th]].

\bibitem{Chen:2010fr}B. Chen, J. Long, J. Zhang,   Phys.\ Rev.\ D {\bf 82}, 104017 (2010), [arXiv:1007.4269v1 [hep-th]].

\bibitem{quasinomral1999}
H.P. Nollert, Class.Quant.Grav.16:R159-R216,(1999). K.~D.~Kokkotas
and B.~G.~Schmidt,
    Living Rev.\ Rel.\  {\bf 2}, 2 (1999)
  [arXiv:gr-qc/9909058]. E.~Berti, V.~Cardoso and A.~O.~Starinets,
  Class.\ Quant.\ Grav.\  {\bf 26}, 163001 (2009)
  [arXiv:0905.2975 [gr-qc]].



\bibitem{Horowitz:1999jd}
  G.~T.~Horowitz and V.~E.~Hubeny,
  Phys.\ Rev.\  D {\bf 62}, 024027 (2000)
  [arXiv:hep-th/9909056].

\bibitem{Briminghan01}
D.Birmingham,
 Phys.\ Rev.\ Lett.\  {\bf 88}:151301,(2002)
 [hep-th/0112055]

\bibitem{Hod:1998vk}
  S.~Hod,
  Phys.\ Rev.\ Lett.\  {\bf 81}, 4293 (1998)
  [arXiv:gr-qc/9812002].

\bibitem{Dreyer:2002vy}
  O.~Dreyer,
  Phys.\ Rev.\ Lett.\  {\bf 90}, 081301 (2003)
  [arXiv:gr-qc/0211076].


\bibitem{BTZ}
M$\acute{a}$ximo Ba$\tilde{n}$ados , Claudio Teitelboim and Jorge Zanelli,
Phys. Rev. Lett.{\bf 69}:1849-1851,(1992)
[hep-th/9204099]

\bibitem{BTZ}
  V.~Cardoso and J.~P.~S.~Lemos,
   Phys.\ Rev.\  D {\bf 63}, 124015 (2001)
  [arXiv:gr-qc/0101052].
  N.~Iizuka, D.~N.~Kabat, G.~Lifschytz, D.~A.~Lowe,
  Phys.\ Rev.\  {\bf D68}, 084021 (2003).
  [hep-th/0306209]. V.~Balasubramanian, T.~S.~Levi,
  Phys.\ Rev.\  {\bf D70}, 106005 (2004).
  [hep-th/0405048]. F.~Denef, S.~A.~Hartnoll, S.~Sachdev,
  Class.\ Quant.\ Grav.\  {\bf 27}, 125001 (2010).
  [arXiv:0908.2657 [hep-th]]. Y.~Kwon, S.~Nam,
  Class.\ Quant.\ Grav.\  {\bf 27}, 125007 (2010).
  [arXiv:1001.5106 [hep-th]].

\bibitem{Crisostomo:2004hj}
  J.~Crisostomo, S.~Lepe, J.~Saavedra,
  Class.\ Quant.\ Grav.\  {\bf 21}, 2801-2810 (2004).
  [hep-th/0402048].

\bibitem{Afshar:2010ii}
 H.~R.~Afshar, M.~Alishahiha and A.~E.~Mosaffa,
  JHEP {\bf 1008}, 081 (2010).

\bibitem{Andy08}D. Anninos, W. Li, M. Padi, W. Song and A.
Strominger, JHEP {\bf 0903}:130,2009, [arXiv:0807.3040].


\bibitem{ChenXu2}B.~Chen and Z.~b.~Xu,
   JHEP {\bf 11}(2009)091
  [arXiv:0908.0057 [hep-th]].

\bibitem{Chen:2009cg}
  B.~Chen, B.~Ning and Z.~b.~Xu,
 JHEP {\bf 1002}:031,2010
  [arXiv:0911.0167 [hep-th]].

\bibitem{Bin10selfdual}
Bin Chen and Bo Ning, Phys. Rev. D{\bf 82}:124027,2010.

\bibitem{Anninos:2010pm}
  D.~Anninos, G.~Compere, S.~de Buyl, S.~Detournay and M.~Guica,
  JHEP {\bf 1011}:119,2010
  [arXiv:1005.4072 [hep-th]].

\bibitem{Son:2002sd}
  D.~T.~Son and A.~O.~Starinets,
    JHEP {\bf 0209}, 042 (2002)
  [arXiv:hep-th/0205051].

\end{thebibliography}
\end{document}